\documentstyle[preprint,eqsecnum,aps]{revtex}
\begin{document}
\tightenlines
\preprint{SU-GP-98/5-2, \quad hep-th/9805207}
\title{A New Approach to String Cosmology}
\author{Gary T. Horowitz}
\address{Physics Department, University of California,
Santa Barbara, California 93106}
\address{Institute for Theoretical Physics, Santa Barbara, California 93106}
\author{Donald Marolf}
\address{Physics Department, Syracuse University, Syracuse, New York 13244}
\date{May, 1998}
\maketitle

\begin{abstract}
We discuss quotients of  Anti-de Sitter (AdS) spacetime by a discrete
group in light of the AdS-CFT correspondence. Some
quotients describe closed universes which expand from zero
volume to a maximum size and then contract. Maldacena's conjecture suggests
that they should be represented in string theory by suitable
  quotients of the boundary
conformal field theory.  We discuss the required identifications, and
construct the states associated with the linearized supergravity modes
in the cosmological background.

\end{abstract}

\input{epsf.tex}
\vfil
\eject

\baselineskip = 16pt
\section{Introduction}

Recently, there has been considerable excitement about a conjecture due to
Maldacena \cite{MAL} (based on earlier work e.g. \cite{GUKL})
 which relates string theory
in Anti-de Sitter (AdS) spacetime to a conformal field theory (CFT) living on
its boundary.
 The conjecture actually applies to all finite energy
excitations about this background, and thus includes all spacetimes which
asymptotically approach AdS.  If correct, this
would provide a nonperturbative definition of string theory for these
boundary conditions. We wish to consider the effect of taking the quotient
of AdS by a discrete subgroup of its isometry group. There are several
motivations for doing so. In three dimensions these quotients include
the BTZ black hole \cite{BTZ} and in higher dimensions there are
analogous black hole solutions \cite{ABHP}.
More importantly, some of these quotients describe
simple cosmological models in which a compact space expands from a
``big bang" and collapses in a ``big crunch". Since
the curvature is locally constant, these 
points of zero spatial volume
are not curvature singularities, but more like conical singularities.
Since isometries of AdS are symmetries of the boundary conformal field theory,
these models should be described in string theory
by an appropriate quotient of
the  original boundary theory.

Applying Maldacena's duality to cosmology will clearly require an
extension of the form of the conjecture given in \cite{GKP,WIT} in which
the boundary at infinity played a crucial role. In particular,
the string theory effective action evaluated on a solution 
with given
asymptotic behavior is believed to be
 the generating function for correlation functions 
in the CFT. The appropriate generalization of this statement
 is not yet clear. For our
purposes, it will suffice to work with the correspondence between states in the
AdS background and states in the CFT first discussed in  \cite{HOOO,WIT}.

This offers a new 
way of investigating string cosmology.
There is an extensive literature applying
string theory to cosmology.  (For a 
recent review and references see \cite{BRU}.)
However most of these discussions are based on the low energy effective action
of string theory, and must make assumptions about what happens when the
curvature reaches the string scale.
(A few notable exceptions are  \cite{BAN,LAWI}.)
In principle, the approach described here would be nonperturbative.
 
In practice, to begin to construct the correspondence we must use a perturbative
approach. The radius of curvature of the AdS spacetime depends on the
product of the string coupling $g$ and the Ramond-Ramond charge $N$. We 
will consider the usual limit where the product $gN \gg 1$ is held fixed 
and $N \rightarrow \infty$. Since $g\rightarrow 0$, Newton's constant
is turned off and supergravity modes do not modify the background cosmology.
At nonzero $g$, the backreaction of these modes should cause the curvature
to grow near the initial and final singularities, producing more realistic
cosmological models.

It is far from clear whether this approach will succeed. Not only
must we assume the validity of Maldacena's conjecture, but the quotient
field theory is highly unusual. In particular,
the quotients we consider are different from the orbifolds discussed
in \cite{KASI} which did not act on the AdS space.
Our quotients
are more analogous to Lorentzian orbifolds \cite{HOST}
obtained by identifying points of Minkowski spacetime under the action of a
discrete boost. 
As shown in \cite{KASI}, the
required quotient of the CFT is somewhat subtle,
and is not just a gauging of the discrete group as in the 
string worldsheet treatment of orbifolds.
In addition, Lorentzian orbifolds are much less well understood
than their Euclidean counterparts, but
there are some indications that the identifications we need will act
rather simply. For example,
the three dimensional black hole is just 
such  a quotient of $AdS_3$, and
we expect it to correspond to
excited states of essentially the same CFT.
In all cases, states in the original theory which are invariant under the
group are included in the quotient theory, and that is all that we will use
below.

It turns out that the quotients required to obtain the cosmology introduce
further complications. For example, the identifications will break all
of the  supersymmetry. Further, we will see that the
symmetry group has a dense set of fixed points on the boundary, so one
cannot simply remove the fixed points and take the quotient of the
resulting space. However, it may be possible to take the quotient of the
quantum states
and operators directly, without trying to realize them as a quantum field theory
on a quotient space.
Given the potential importance, it seems
worthwhile 
to try to  construct this theory. Here we take the first steps in
this direction. We show how to construct the states associated with linearized
supergravity modes on the cosmological background. Interactions and
winding sectors will require further investigation.

Another motivation for constructing this cosmological solution
is that it might help construct a background independent formulation of
string theory. As emphasized by Banks \cite{BAN}, it is only in the context of
a closed universe that all moduli can fluctuate, since it does not
require infinite energy to excite them. This may thus be
the appropriate starting point for trying to understand why the
compact dimensions take the form that they do. 

In the next section we review some of the spacetimes that can be
constructed by taking quotients of AdS. For the $AdS_3$ case, we
briefly describe the corresponding states in the CFT. In \cite{MAST}
it was argued that the BTZ black hole is naturally associated with
a density matrix in the CFT. We will see that this is not the case
for more general black holes.
In section III we discuss the cosmological models and construct
the states associated with linearized supergravity modes. Section 
IV contains a brief discussion.

\section{Quotients of anti de Sitter space}

We begin by considering some of the spacetimes that can be constructed
by taking quotients of AdS space. Since this space arises in string theory
with constant dilaton  and a Ramond-Ramond field proportional to the
volume form, all quotients are also classical solutions of  string theory. 
 These include the BTZ black hole
\cite{BTZ}
and its higher dimensional generalizations \cite{ABHP}, 
the `wormhole' solutions of \cite{ABBHP,ABH}, and various cosmological
solutions whose spatial sections are compact manifolds of constant negative
curvature.
The cosmological models will be discussed in more detail in
section \ref{compact}.

As is well known, $n+1$ dimensional AdS space can be obtained by taking
the surface\footnote{For simplicity, we will consider the case of
 unit radius here.} 
\begin{equation}
\label{embed}
-T_1^2 + \sum_{i=1}^n X_i^2 - T_2^2 = - 1
\end{equation} 
in a flat spacetime of signature $(n,2)$
and Cartesian coordinates $(X_i,T_1,T_2)$. 
A convenient parameterization for discussing quotients is to set 
$T_2 = \sin \tau$, so that a constant $\tau$ surface is a
 constant negative curvature hyperboloid
of radius $\cos \tau$. The resulting metric takes the Robertson-Walker form
\begin{equation}
\label{invmet}
ds^2 = - d\tau^2 + \cos^2 \tau \ d \sigma^2_n
\end{equation}
where $d\sigma^2_n$ is the metric on the (unit) 
hyperboloid. Worldlines which remain at a fixed point on the hyperboloid
are timelike geodesics.  In this form
of the metric, spatial symmetries of the hyperboloid are spacetime isometries.
(This is not the case for e.g. the globally static form of the metric which
also has spacelike surfaces with  constant negative curvature.)
Such symmetries form
an $SO(n,1)$ subgroup
of the full $SO(n,2)$ symmetry group of AdS
which acts within surfaces of constant $\tau$.  For the case
$n=2$, this subgroup acts diagonally with respect to the local decomposition
$SO(2,2) \sim SO(2,1) \otimes SO(2,1)$.  
Note that the action of
such an isometry on the entire spacetime (covered by our coordinate system)
follows from its action on the spatial slice at $\tau=0$.
The representation (\ref{invmet}) is convenient for the construction of 
spacetimes with a moment of time symmetry, 
to which we will confine ourselves in this paper.
However, we expect that 
the `spinning' cases of \cite{BTZ,ABH} can be treated in much the
same way.

The coordinates used in (\ref{invmet}) do not cover all of AdS space, but only
the domain of dependence of the $\tau=0$ surface. 
 Let us denote this region by $R$.
If we choose a conformal compactification
of (the covering space of)
AdS space so that its boundary is a timelike cylinder, then
$R$ is the interior of the cone shown below.
The entire null cone corresponds to the coordinate singularities
at $\tau = \pm \pi/2$. 

\medskip
\centerline{ \epsfbox{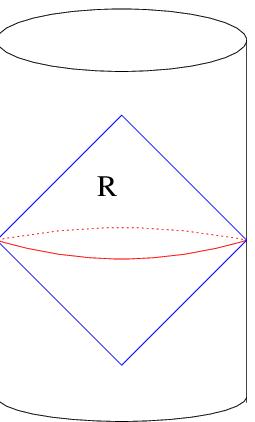}}

\medskip
\centerline{Fig. 1.  The region $R$ covered by (2.2) and
the conformal compactification of AdS.}

\medskip

It will be useful to introduce on the boundary
a time coordinate $t$ and angular coordinates
for the $n-1$ sphere.  We choose our conformal
compactification so that the boundary metric is
\begin{equation}
\label{bdymet}
ds^2 = - dt^2 + d\theta^2 + \sin^2 \theta\ d\Omega^2_{n-2}
\end{equation}
where $d\Omega_{n-2}^2$ is the metric on the unit $n-2$ sphere.
We take the intersection of our cone with the boundary to be the 
surface $t=0$.  To describe the symmetries of the hyperboloid, it will be
convenient to think of it as embedded in
an auxiliary Minkowski spacetime via 
(\ref{embed}) with $T_2 = 0$. In terms of this embedding, the $t=0$
sphere can be thought of as the sphere of null
directions which naturally forms the boundary of
the hyperboloid.  
In this way, we may 
choose angular coordinates on the boundary that are adapted to our choice
of coordinates $T_1,X_i$.  In particular, we will take the polar angle $\theta$
to have its singularities (the north and south poles) on the
$X_1$ axis.

Since the metric (\ref{invmet}) is invariant under the full $SO(n,1)$ group, 
we may take quotients of this spacetime under discrete subgroups $\Gamma$ of 
$SO(n,1)$ so long as no conical singularities are produced (or perhaps, as
in \cite{ABBHP,ABH}, so long as such singularities are hidden behind horizons).
Often, the resulting quotient space can be extended beyond the region $R$
shown in Fig 1.  This happens whenever
the quotient operations produce no singularities on a region of AdS
space that is larger than $R$.   However, we can still use (\ref{invmet}) 
to study
such compactifications since the extensions are 
essentially unique.  The point is that, if we impose asymptotically AdS
boundary conditions at infinity when appropriate, the surface $\tau=0$ provides
initial data for the entire covering space of AdS.  Thus, the action of a 
Killing field near this surface defines its action everywhere.

Perhaps the simplest example is when $\Gamma$ is the cyclic group 
generated by a single boost.  We will take the boost
which generates $\Gamma$ to be in the $T_1,X_1$ plane of (\ref{embed}).
This boost has no fixed points in the interior of the $\tau =0$ slice
so the interior spacetime is straightforward to construct.
It is just the BTZ black hole or one
of its higher dimensional generalizations.
We are, however, more interested in what happens to the boundary
under this boost.  Note that any symmetry of the interior will
induce a conformal symmetry on the boundary.  By appropriately choosing
the conformal factor of the boundary metric, we may in fact take
this to be a Killing symmetry of the boundary.  Thus, we may also use
$\Gamma$ to obtain the boundary of the new spacetime directly 
as a quotient of the original AdS boundary.  This will allow us to 
see how states in the relevant CFT's are related.  

Note that our boost has
two fixed points on the boundary of the $\tau =0$ slice, 
corresponding to the two null vectors
in the $T_1,X_1$ plane. One is attractive and the other is repulsive.
These are at the poles of our angular
coordinate $\theta \in [0,\pi]$.
If the boost has boost parameter $\lambda$, then
the action of the boost on the boundary is $\theta \rightarrow \tilde   \theta$
where
\begin{eqnarray}
\label{bact}
\cos \tilde \theta = {{\sinh \lambda + \cosh \lambda \  \cos \theta} \over
{\cosh \lambda + \sinh \lambda \ \cos \theta}}
\end{eqnarray}
and none of the other angles are altered\footnote{Since $\cos \theta$ is
the $X_1$ component of the velocity (that is, $dX_1/dT_1$)
of the null rays associated
with that value of $\theta$ in our auxiliary
Minkowski space, the form of this 
transformation follows readily from the usual action of a
boost on velocities in flat spacetime.}.

Let us excise the fixed points from the $t=0$ sphere at infinity and consider
the domain of dependence ${\cal D}$
in the boundary of the resulting set (topologically, ${\cal D}$ is
${\bf R} \times S^{n-2} \times {\bf R}$).  The
boost  has no fixed points in this domain, so the quotient of this
domain by $\Gamma$ is well defined and is topologically $S^1 \times
S^{n-2} \times {\bf R}$.    
To determine the conformal
structure on this space, 
let us introduce null coordinates $u$ and $v$ (which take
values in $(-\pi/2,\pi/2)$ 
) in the domain
${\cal D}$ through $u = t - \theta + \pi/2 $ and $ v = t + \theta - \pi/2 $.
 We will also rescale the boundary metric (which we are always free to do)
by $\sin^{-2} \theta$.  This will yield a metric well adapted to use
with $\Gamma$. The metric on ${\cal D}$ is then
\begin{equation}
ds^2 =- {{du dv} \over {\cos^2 \left({{v-u}\over 2}\right)}} + 
d\Omega_{n-2}.
\end{equation}

The action of the identifications on both $u$ and $v$ can be read off
from (\ref{bact}), since the $t=0$ circle is given by $- u=v=\theta-\pi/2$.
This can be simplified by realizing that (\ref{bact}) is generated by the
vector field $-\sin \theta {\partial \over {\partial \theta}}$ and
thus that the action on $u,v$ is generated by 
$\cos u {\partial \over {\partial u}} - \cos v {\partial \over {\partial v}}$.
 In terms of the usual Virasoro
generators this corresponds to $L_1 +L_{-1} - (\bar L_1 + \bar L_{-1})$.
 If we introduce new null
coordinates 
$\alpha = \int_0^u {{du} \over {\cos u}}$,
$\beta = \int_0^v {{dv} \over {\cos v}}$, then this vector field
may be written
${\partial \over {\partial \alpha}} -{\partial \over {\partial \beta}}$.
  Thus, the identifications are just
$(\alpha, \beta) \rightarrow (\alpha + \lambda, \beta - \lambda)$ and
the appropriate conformal structure on   $S^1 \times S^{n-2} \times {\bf R}$
is 

\begin{equation}
\label{cf}
ds^2 =- {{\cos u \cos v} \over {\cos^2 \left({{v-u}\over 2}\right) }} \
d\alpha d\beta +  d\Omega_{n-2}.
\end{equation}
One can show that
$\left({\partial \over {\partial \alpha}} -
{\partial \over {\partial \beta}} \right)\left[ \cos u \cos v \cos^{-2}
\left({{v-u}\over 2}\right)\right] = 0$ 
so the identifications yield a smooth metric for any value of $\lambda$.

We now consider the case $n=2$ corresponding to $AdS_3$. The above
identifications yield
 the BTZ black
hole with a nonzero mass (which depends on the actual period chosen).
 Since $S^{n-2} = S^0$ consists of two points, the boundary at
infinity is two copies of the cylinder $S^1\times {\bf R}$,
corresponding to the asymptotic 
regions on both sides of the black hole.
The identification in the interior can be made explicit by writing
$d\sigma^2_2 = dz^2 + \cosh^2 z\ dw^2$. 
The boost corresponds to translations
of $w$.\footnote{The zero mass black hole  requires a different identification
which corresponds to making $y$ periodic in the metric
$d\sigma_2^2 = x^{-2}(dx^2 + dy^2)$.
This symmetry has only a single fixed point at infinity, and so the boundary
is a single copy of $S^1\times {\bf R}$.} The form of the metric (\ref{invmet})
for the
BTZ black hole is perhaps unfamiliar, since the coordinates are  adapted to 
timelike geodesics as opposed to the time translation symmetry.

Suppose that Maldacena's conjecture is correct and that string theory
on $AdS_3$ (times a compact space)  is described by a boundary
conformal field theory.  Then, since the above spacetime 
 can be obtained from $AdS_3$
by taking a quotient with respect to $\Gamma$, it seems reasonable
to expect that the quotient theory on the boundary 
includes the same conformal field theory
defined on the quotient space 
(although it may include twisted sectors as well).

If we were to start 
directly with the field theory on $S^1 \times S^0 \times {\bf R}$, 
it would be natural to use the 
null vector fields ${\partial \over {\partial \alpha}}$ and
${\partial \over {\partial \beta}}$ (which are conformal
Killing fields of (\ref{cf}) when $n=2$) to pick out preferred sets
of positive frequency modes and ask them to annihilate the
vacuum state.
This would give the usual
vacuum on $S^1 \times S^0 \times {\bf R}$. 
However, we have constructed the conformal field theory by
taking a quotient of the boundary spacetime manifold under the group 
$\Gamma$.    Recall that the vacuum $|0\rangle$ of the original
conformal field theory is
in fact invariant under the entire AdS group, and so is, in particular, 
invariant under the group $\Gamma$. Since $|0\rangle$ represents AdS
before the identification, it is natural to assume \cite{MAST}
that 
it describes the BTZ black hole in the quotient.
However, the vacuum $|0\rangle$ 
will contain correlations between the two halves of the original $t=0$ boundary
circle in exactly the same way that the Minkowski vacuum contains correlations
between the left and right Rindler vacua.  As a result, the image
of the vacuum $|0\rangle$ under the quotient operation cannot be the
usual vacuum of the conformal field theory on $S^1 \times S^0 \times {\bf R}$ 
(which does not contain such correlations).  Instead, it must be a state
that, from 
the perspective of
one asymptotic region, 
is a mixed state 
and has correlations with the other asymptotic region. 
In fact, this state
would appear thermal, since the timelike vector field
${\partial \over {\partial \alpha}}+
{\partial \over {\partial \beta}}$, when restricted to one Rindler wedge,
acts like a Rindler time translation in the covering spacetime. 
This is very similar
to the observation in \cite{MAST} that the black hole corresponds to a 
thermal state in the CFT. The relation between our description and theirs
is simply that \cite{MAST}
viewed infinity as a copy of two dimensional Minkowski spacetime,
whereas the above discussion uses the conformal compactification of
Minkowski spacetime in which space is a circle. The domain ${\cal D}$
is just the left and right Rindler wedges.

Although it may seem appealing that the black hole is described by a 
thermal state in the CFT, this does not always seem to be the case.
 The density matrix was a result of the
fact that the boundary of the quotient spacetime is disconnected and that
the state was studied from only a single asymptotic region.
In higher dimensions, this space remains a connected manifold.
The quotient spacetime has only a single asymptotic region, though the
topology of this new region is now different from that of the original
infinity. The original vacuum state is still conformally invariant, so
it should be defined in the quotient theory. But now it will be a pure state.
The structure of the vacuum state in these cases
is worthy of further investigation, however this is complicated by the
fact that the boundary no longer has a time translation symmetry.  

One can also construct examples in $2+1$ dimensions of black holes
with a single asymptotic region. These 
are the wormhole spacetimes of
\cite{ABBHP}.  
For such cases, the spatial sections have the topology of a punctured
Riemann surface.
The simplest versions are formed by taking the
quotient under a subgroup $\Gamma$ generated by two boosts.  The group
$\Gamma$ can best be described by displaying a fundamental domain
for the $\tau =0$ hyperboloid:

\centerline{\epsfbox{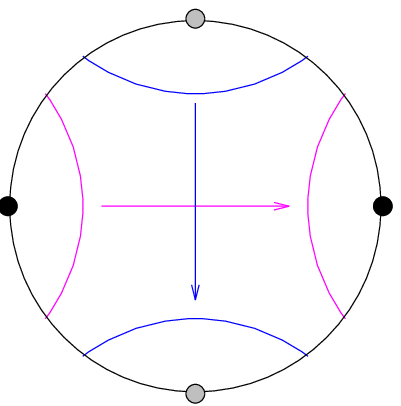}}
 
\medskip
\centerline{Fig. 2. A fundamental domain and identifications
for a simple wormhole spacetime.}
\medskip

The boosts identify the sides of the fundamental region as indicated
by the arrows, and each boost
has a set of two fixed points (shown above, lightly shaded for one
boost and heavily shaded for the other) as before.
As discussed in \cite{ABHP,ABH} the entire group has an infinite
number of fixed points.  Nevertheless, these fixed points are not
dense and a fundamental domain on the boundary
can be identified.  Thus, the fixed points may be excised and the 
quotient is readily constructed.  
The resulting spacetime
is again a black hole with a  single asymptotically AdS region.
As before, one might expect that string theory in this background is
described (at least 
in part) by the original CFT on the quotient of the boundary.
The unexcited wormhole should correspond to the
image of the original vacuum $|0\rangle$. As mentioned above,
 since there is only one asymptotic
region,  the image of the original vacuum $|0\rangle$ should now be a
pure state and not a mixed state.
However, this state
contains correlations between regions that were close to the 
excised points before
taking the quotient.
Thus, we expect the wormhole spacetime to correspond to
some state with these extra correlations.  
The relation between the BTZ and wormhole boundary states should be
similar to the relation between the Hartle-Hawking
vacuum on the Kruskal spacetime and the vacuum studied in \cite{LM}
on the ${\bf RP}^3$ geon\cite{RP3} with a component of the boundary
in the present case playing the role of an entire exterior Schwarzschild
region in \cite{LM}. 
It is an interesting question if
the wormhole state has finite energy with respect to the natural 
vacuum on $S^1\times {\bf R}$ or if it lies in a different superselected
sector of the field theory.
 The answer is likely to shed light on the description of topology
change in terms of the conformal field theory.  Higher dimensional versions
of this spacetime may also be of interest.

\section{Cosmological models}
\label{compact}

\subsection{Spatially compact quotients}

 From our point of view, the most interesting spacetimes that can be constructed
as quotients
of anti-de Sitter space are spatially compact.  These follow by compactifying
the $\tau =0$ hyperboloid using an appropriate discrete group $\Gamma$.
By choosing $\Gamma$ 
properly,
we can obtain cosmological models
with any constant negative curvature space as its spatial section.
For definiteness, we will concentrate on the case of $AdS_3$, where 
this is just the construction of the $g \ge 2$ Riemann surfaces by
taking quotients of the two dimensional hyperboloid.  We refer the reader to
 \cite{ABBHP,ABH,BV}
for details, but give a brief summary below.
The genus $g$ surface is the quotient of the hyperboloid under a group
$\Gamma$ which is generated by $2g$ generators $\{\gamma_1,...,\gamma_{2g}\}$.
The first of these ($\gamma_1$) is a boost in, say, the $T_1,X_1$ plane and the 
others are just conjugations of $\gamma_1$ by $\pi/2g$
rotations about the $T_1$-axis.  Thus, a
fundamental region for $\Gamma$ is an equilateral $4g$-gon centered
at $X_i=0$.  This is shown below for the case $g=2$.

\centerline{\epsfbox{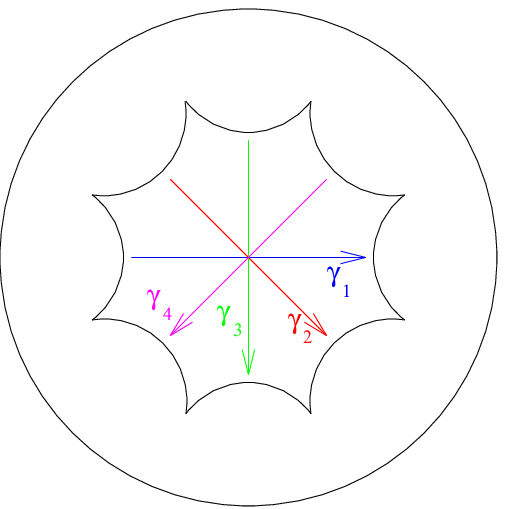}}

\medskip
\centerline{Fig. 3. The fundamental domain $R_0$ and the
identifications for $g=2$.}
\medskip

\noindent
Copies of the fundamental region tessellate the hyperboloid and are in 
one-to-one correspondence with the elements of $\Gamma$.  It will
be convenient to refer to the fundamental region at the origin
as $R_0$ and the image of this region under $\gamma$ as $R_\gamma$.

The magnitude of the boosts (they are all the same) is uniquely 
determined by the requirement that the quotient possesses
no conical singularities
or, equivalently, that the generators satisfy the relation:

\begin{equation}
\gamma_4 ... \gamma_3^{-1}  \gamma_2 \gamma_1^{-1}  \gamma_{2g}^{-1}
 ... \gamma_3  \gamma_2^{-1}  \gamma_1 = \openone.
\end{equation}
See \cite{BV} for a discussion of how this construction is related to
the more familiar homotopy generators satisfying $a_1 b_1 a_1^{-1} b_1^{-1}
...a_g b_g a_g^{-1} b_g^{-1} = \openone$.
For the case $g=2$, the associated boost parameter is $\lambda =
\ln(\sqrt 2+1+(2 \sqrt 2 +2)^{1 \over 2}) $.

What is the action of this group on the boundary at infinity? 
 The action of the generators
on the $S^1$ at $t=0$ follows directly from (\ref{bact}).
A generic element of $\Gamma$ is, however, a bit more complicated.
 From \cite{BV}, we know that any such element 
is again a boost, 
but in general it will be associated with a plane in the auxiliary
Minkowski space (eq. \ref{embed}\ with $T_2 = 0$) that does not contain
the $T_1$-axis.  If we regard the $\tau=0$ hyperboloid
as the Poincar\'e disk, then
these boosts act as rather general hyperbolic M\"obius transformations.  Such
transformations still have exactly two fixed points on the boundary, one
attractive and one repulsive, though
these fixed points can now be arbitrarily close together.  In fact, it 
follows from the ergodicity results of \cite{BV} (see appendix
\ref{dense}) that the
set $S$ of points on $S^1$ which remain fixed under some element of $\Gamma$
is dense in $S^1$.
Thus, we can no longer excise them and identify a nice
fundamental region with which to work.  Another property of the
action of $\Gamma$ that is perhaps even more unpleasant is that, since
any fixed point of $\gamma \in \Gamma$ on $S^1$ is in fact  an {\it attractive}
fixed point either for
the cyclic semigroup generated by $\gamma$ or the one generated
by $\gamma^{-1}$, the set
of such attractive fixed points
in $\Gamma$ is also dense in $S^1$.  As a result, given
any open set $U$ and any point $x$ in $S^1$, there is some element of
$\Gamma$ that maps $x$ into $U$. 

The action of $\Gamma$ on the boundary away from the $t=0$ circle
is only slightly better.
To see this, note that we 
can deduce the action of $\Gamma$
on the entire boundary directly from (\ref{bact}) by
introducing null coordinates $u = t - \theta$ and $v = t+ \theta $ on 
the boundary.  Both of these are periodic with period $2\pi$ in the sense
that $u$ and $u+2 \pi$ label the same null ray, and similarly for $v$.
The circle $t=0$ is just the circle $v= -u = \theta$.  Since each element
of $\Gamma$ is a symmetry of AdS space, it acts as a conformal transformation
on the boundary, mapping rightmoving null rays to rightmoving null rays.  
In other words, any element of $\Gamma$ will act on the boundary
in the form $(u,v) \rightarrow (f(u),g(v))$.  Thus, we can read off the 
functions $f,g$ from (\ref{bact}) and we see that they are identical (up
to signs).
This means that the set of points that are fixed by some element of
$\Gamma$ is still dense in the boundary, but (since $f(\theta)=-g(-\theta)$ 
as functions from
$S^1$ to $S^1$)
that all of the attractive
fixed points of cyclic 
semigroups
lie in the $t=0$ circle $v=-u$.  Thus, it is again impossible to 
identify a fundamental region and construct a quotient 
manifold.  It would be of interest to characterize the topological
space that results from removing the fixed points and then performing the
quotient.  It may well be Hausdorff, but it is certainly not a smooth
manifold.

\subsection{Constructing the States}

It is clear from the above discussion that, for the cosmological models,
one cannot take the quotient of the 
space 
in which the original
boundary conformal field theory
lives.  What we would like to do instead is to take the quotient
of the original conformal field theory (that is, the collection of 
states and operators) {\it directly}, constructing a `quotient
state space' and a `quotient operator algebra' without worrying about
whether or not this quotient theory can be thought of as a quantum field
theory on some spacetime manifold.

Before explaining how this can be (partially) 
achieved, let us quickly review the correspondence  that has been established
between states
in the uncompactified AdS and the CFT.
We will focus on the case where the background spacetime is $AdS_3 \times
S^3 \times T^4$. The
boundary field theory is a two dimensional, ${\cal N} = (4,4)$,
superconformal field theory. 
Since the isometry group of $AdS_3$ is $SL(2,R) \times SL(2,R)$, one can
define isometries $L_1,\ L_0,\ L_{-1}$ and
$\bar L_1,\ \bar L_0,\ \bar L_{-1}$ with the 
commutation relations of the Virasoro algebra (see \cite{MAST,BKL} 
for the explicit
expressions and further discussion of the modes). The function $\phi_{0}$
 satisfying $L_1 \phi_0 = \bar L_1 \phi_0 = 0$,
$L_0 \phi_0 = \phi_0 = \bar L_0 \phi_0$ and which vanishes at infinity
is a positive frequency
solution to the massless wave equation $\nabla^2 \phi = 0$
on $AdS_3$ and can be viewed as a primary state. There is a corresponding 
chiral primary state $|h\rangle$
in the CFT. All other one particle states of this field in $AdS_3$
are of the form $ L_{-1}^m \ \bar L_{-1}^n \phi_0$,
and the corresponding CFT states are of the form 
 $L_{-1}^m \ \bar L_{-1}^n |h\rangle $.
In the CFT these are all essentially BPS in that their mass does not receive
quantum corrections. 
Similar statements apply to all the supergravity modes including the massive
Kaluza-Klein states resulting from the compactification on $S^3 \times T^4$
\cite{MAST,BOE}.

Roughly speaking, to obtain the state associated with a 
linearized supergravity mode in the cosmological background, we would like
to proceed as follows.
Given a supergravity mode in the cosmology, lift it 
up to the original uncompactified AdS spacetime. The result is a periodic
function.
For each domain $R_\gamma$, consider a new function which is
zero outside $R_\gamma$ but agrees with the original function inside.
This new function is now square integrable and 
corresponds to a
state in the CFT. The cosmological mode is then associated with the sum (over
all domains $R_\gamma$) of these
states. The problem, of course, is that this sum does not
converge to a normalizable state in the Hilbert space.
Below we explain how this problem can be resolved.

Our problem is similar to difficulties that have arisen in other contexts 
\cite{AH,KL,DM1,ALMMT,DM2} where one wished to take a quotient of 
a quantum theory with respect to a noncompact group, and we will borrow
the techniques used there.  Such techniques (which we refer to as ``induction," 
though the names in the literature vary) are in turn based on the rigged
Hilbert space methods of \cite{Gel} used in constructing generalized
eigenstates of operators with continuous spectra.
We will not review the details here, but merely apply such methods 
below\footnote{We will, however, insert and remove various complex
conjugations relative to \cite{ALMMT,DM2} to make the connection between
the various theories more explicit.}.
We encourage the reader to consult the above references (especially
section II of ref. \cite{ALMMT} and ref. \cite{DM2}) 
for further details and a discussion
of the general approach.

Let $C$ denote the original conformal field theory, and let $Q$ denote the
quotient that we wish to construct.
The method involves finding `distributional states' of $C$  which are
invariant under $\Gamma$.  This is the same idea as saying that
certain (non-normalizable) Fourier modes on the real line are
invariant under discrete translations.  It turns out that, 
because of the complicated action of $\Gamma$ on the spacetime on which
$C$ lives, it is difficult to 
control
such a space of distributions by
working with $C$ alone.  Instead, as outlined above, we will first
show that the linearized supergravity states on the cosmology can be
thought of as distributional states in the theory on $AdS_3 \times S^3 
\times T^4$ and that the induction techniques define
the proper Hilbert space structure on these distributions.  We can then
use the correspondence of \cite{MAST,BOE} to carry this over to the CFT
and thus define certain states in $Q$ as distributional states in $C$.  These
will be the states in $Q$ that correspond to the linearized
supergravity states on the cosmology.

Recall that the space of one particle states of linearized supergravity on
$AdS_3 \times S^3 \times T^4$ can be associated
with an $L^2$ space on the $\tau=0$ hyperboloid ($\times S^3 \times T^4$), 
defined using the volume
element $dv$.  These states are analogues of the Newton-Wigner states
in flat spacetime.  That is, a given function $f \in L^2$ does not
represent the state created by the quantum field smeared directly with
the test field $f$, but instead this correspondence involves the action
of $\sqrt{\nabla_0^2}$, where $\nabla_0^2$ is the Laplacian on
the $\tau=0$ surface.  

In order to define what is meant by ``distributional states,'' induction
requires the choice of a dense subspace $\Phi$ of the state space.
To introduce our choice of $\Phi$, consider 
polar coordinates $(r,\theta)$ on the hyperboloid such that the metric
becomes
\begin{equation}
d\sigma^2 = dr^2 + \sinh^2 r \ d\theta^2.
\end{equation}
Let $\Phi$
be the space of one particle states associated as above with
smooth $L^2$ functions which {\it A}) vanish at infinity
at least as rapidly as $e^{-r}/r^{1 + \epsilon}$
and {\it B}) have
vanishing integral over the $\tau=0$ surface.  Note that 
a topology on $\Phi$ is provided by its inclusion in the original
Hilbert space.  

To each state $\phi \in \Phi$, we will associate a function $\eta(\phi)$
on the $\tau=0$ surface that is invariant under the action of $\Gamma$.
To do so, recall
that the elements $\gamma$ of $\Gamma$ are in one-to-one correspondence
with the images $R_\gamma$
of the fundamental region that tessellate the hyperboloid.
Since all such images have the same area, the number of 
images located near the coordinate value $r$ is roughly $e^r$ for large
$r$.  It follows that the sum

\begin{equation}
\label{imsum}
[\eta({\phi})](x) \equiv \sum_{\gamma \in \Gamma} \phi(\gamma(x))
\end{equation}
converges absolutely at each point to define a function that
is invariant under the action of $\Gamma$. 

In fact, given any
two functions $\phi_1,\phi_2 \in \Phi$, the integral 
\begin{equation}
\label{int}
\int \ dv \ 
[\eta(\phi_1)](x) \phi_2(x)
\end{equation}
converges absolutely.  Thus, we see that 
$\eta$ defines a map from $\Phi$
to its topological dual $\Phi'$ where the action of $\eta(\phi_1)$ on
$\phi_2$ is given by (\ref{int}).

It is from the image of $\eta$ in $\Phi'$ that we will construct states
of the quotient theory.
In fact, we let any $\eta (\phi)$ belong to our quotient Hilbert
space.  Using $*$ to denote complex conjugation, we
introduce an inner 
product
on this space as follows: If
$\overline{\phi}_1 = \eta (\phi_1)$ and $\overline{\phi}_2 = \eta(\phi_2)$, 
we set

\begin{equation}
\label{ip}
\langle \overline{\phi}_1, \overline{\phi}_2 \rangle
= \overline{\phi}_2 (\phi_1^* ) = \int \ dv \ \overline{\phi}_2(x) \phi_1^*(x)
= \int \ dv \ \phi_2(x) \overline{\phi_1^*}(x).
\end{equation} 

In general 
\cite{ALMMT,DM2}, this construction has the property that any operator
on the original Hilbert space which preserves
the space $\Phi$ will induce a corresponding operator in the quotient
theory and that, with the inner product (\ref{ip})
the *-algebra of all such operators will be preserved. 
The product is appropriately symmetric and bilinear and it 
is positive definite 
on the image of $\eta$, so that it does indeed define a Hilbert space.
The kernel of $\eta$ is not a problem here.  Although a given
distribution $\overline{\phi}_1$ in the image of $\eta$ 
will not be associated with a unique
element $\phi_1$ of $\Phi$, we see that (\ref{ip}) is independent
of which $\phi_1$ is chosen.  All states
in the kernel of $\eta$ are mapped to the zero distribution
in $\Phi'$ and thus to the zero vector in the resulting
Hilbert space.  

As stated above, the functions $\eta(\phi)$ are invariant under $\Gamma$.
We would therefore like to think of them as representing the
linearized supergravity states on 
the cosmology.  However, for this to work it is important that (\ref{ip}) 
agree with  the inner product in 
this sector of the cosmological string theory.  Since we may write
(\ref{ip}) as 

\begin{equation}
\label{int2}
\langle \overline{\phi}_1, \overline{\phi}_2 \rangle
= \sum_\gamma \int_{R_\gamma} \ dv \ \overline{
\phi}_2 \phi_1^* =
\int_{R_0} \ dv \ \overline{\phi}_2 \sum_\gamma (\gamma \phi_1^*) 
= \int_{R_0} \ dv \ \overline{\phi}_2 (\overline{\phi}_1)^*,
\end{equation}
we see that this is the case.

Finally, we note that this construction in fact yields all one particle
linearized supergravity states (and no 
extra states) on the cosmology\footnote{Since the cosmology is time dependent,
particle number is not conserved. The states we have constructed can be 
viewed as one particle states at time $\tau = 0$.}.
This is equivalent to the statement that the image of $\eta$
consists exactly of
those smooth functions 
which are invariant
under $\Gamma$ 
and which are orthogonal to the constant
function in the $L^2$ inner product on the compactified space.
To see 
that this is so, consider any smooth real function 
$\rho_0 \ge 0$ of compact support that does not vanish anywhere on $R_0$.
Now, define 
\begin{equation}
\rho = {{\rho_0} \over {\sum_{\gamma \in \Gamma} \rho_0 }}.
\end{equation} 
If we would extend the definition of $\eta$ to functions whose integral
over the hyperboloid did not vanish, we would have $\eta(\rho) =1$.
It follows that, given any function $f$ on the hyperboloid which is
{\it A}) invariant under $\Gamma$ and {\it B}) orthogonal to the constant
function in the inner product on the compactified space (so that, in 
particular, its integral over $R_0$ vanishes), the 
function 
$(f \rho)(x) = f(x) \rho(x)$
lies in $\Phi$ and satisfies $\eta(f \rho) =f$.   The
fact that $f \rho$ has zero total integral follows from
(\ref{int2}) above with $\phi_1=(f \rho)^*$ and $\phi_2 = 1$.  It
is also clear that any $\phi \in \Phi$ satisfies $\int_{R_0}
\overline{\phi} =0$ so that we have not defined any ``extra'' states.
Thus, the image of $\eta$ is exactly the Hilbert space of
one particle linearized supergravity states on the cosmology.

In this way, we have written a certain subspace 
of the cosmological string theory Hilbert space 
in terms of a subspace of the uncompactified string theory. 
It is now straightforward to
use the correspondence of \cite{MAST,BOE} to carry this construction over 
to the conformal field theory $C$.  Thus, we have an associated
space of states $\Phi_C$ and an associated map $\eta_C: \Phi_C
\rightarrow \Phi_C'$.  Induction then produces 
a small Hilbert space $Q_0$
in terms of the distributional states in 
the image of $\eta_C$
and we can see that 
such states correspond to the linearized supergravity states on
the cosmological background.  If Maldacena's conjecture is correct, 
we expect this small Hilbert
space to be part of a larger theory $Q$ which is equivalent to
the entire cosmological string theory.

A final comment is in order 
concerning the sum (\ref{imsum}).
Recall that the correspondence between states in the linearized
supergravity theory and the conformal field theory states holds, 
for a particular
state, only in the limit $g \rightarrow 0$.  The reader may therefore
wonder about our infinite sum (\ref{imsum}) and the fact
that our argument implicitly involves interchanging the $g \rightarrow
0$ limit with this sum.  This may to some extent be justified
by noting that, for finite $g$, given any anti de Sitter invariant measure
of the accuracy to which the correspondence in \cite{MAST,BOE} holds
for a given mode $f$, this correspondence must hold to the same accuracy
for each of the images $\gamma f$ of this mode.  This observation provides
a certain uniformity of the $g \rightarrow 0$ limit with respect to our
series (\ref{imsum}).

\section{Discussion}
\label{disc}

We have seen that one can 
take the quotient of 
anti de Sitter spacetime and obtain
a simple cosmological model, with compact spatial sections that expand from
zero volume to a maximum size and then contract back to zero volume. 
Using the AdS-CFT correspondence 
we have shown how
to construct a Hilbert space out of distributional
states in the CFT 
which are 
invariant under the
discrete group and hence live in the quotient. These states should describe
linearized supergravity modes in the cosmological background. Although
we have focused on the three dimensional case, 
 a similar construction will
be valid in higher dimensions as well.

This is clearly only the first small step toward constructing a 
quantum theory $Q$ which might provide a nonperturbative definition of
string theory for cosmology.  We have not treated operators 
in $Q$ that would correspond to the zero modes of the supergravity theory on the
compact space, discussed interactions, or possible winding sectors.
With regard to the latter,
 one might argue that near the moment of maximum expansion, all winding
modes will be rather heavy, with masses of order the radius of curvature of AdS.
However, near the singularities, one would expect the winding
modes to be very important.

We suspect that the final theory will be
quite unusual: from our discussion of the
action of $\Gamma$ on the spacetime associated with the original conformal
field theory $C$, 
we do not expect the theory $Q$ to be a quantum field theory on any
smooth spacetime manifold.  The hope is that, by relating it to
the theory $C$ through, for example (\ref{imsum}), 
the theory $Q$ can nonetheless be sufficiently well controlled.

Let us, for the moment, suppose that this theory can be constructed.
The payoff would be enormous.  The resulting theory
 would then provide a nonperturbative description of string cosmology.
In the happy event that the theory can be defined without reference to $C$, it
would likely be background independent (since the asymptotic AdS boundary
conditions which provided the main background dependence before have
been eliminated).
 What  is the
range of states that could be described?  At the very least, one would 
expect the answer to be ``all states associated with a spacetime
of the given topology.'' 

However, topological fluctuations might also be allowed.
Indeed, any genus greater than two
Riemann surface can be obtained as a quotient of the $\tau =0$
hyperboloid. So, 
at the linearized supergravity level, the associated
quotient theories $Q$ all fit together inside the original conformal
field theory $C$.  If the interactions
do not cleanly separate in the same way, but instead
mix the sectors of $C$ that we have associated with different
spatially compact manifolds, then 
it appears natural to associate
such behavior with a description of topology change. 
These are clearly interesting
issues for future investigation.

\vskip .5cm

{\bf Note Added:} We have recently learned that the results derived
in the appendix were previously published in \cite{extra}. 

\acknowledgements 
We thank I. Bengtsson, C. Rovelli, and the participants of the duality
program at the Institute for Theoretical Physics, Santa Barbara 
for helpful discussions.
This work  was supported in part by NSF grants PHY94-07194, PHY95-07065,
PHY-9722362, and funds provided by Syracuse
University.

\appendix

\section{The fixed points of $\Gamma$ are dense for any 
genus $\ge 2$ surface.}
\label{dense}

As stated in section \ref{disc}, these results have appeared
previously in \cite{extra}.
\medskip

We first prove a lemma concerning the group $\Gamma$ 
that compactifies the hyperboloid to any $g \ge 2$ surface and a lemma
concerning the associated action of $\Gamma$ on $S^1$. The main
theorem then follows.

\medskip

{\bf Lemma 1:  Let $\lambda_\gamma$ be the boost parameter associated
with the boost $\gamma \in \Gamma$.  The set $\{ \lambda_\gamma :
\gamma \in \Gamma\}$ is bounded below.}

\medskip

Proof:  First note that, 
on our Riemann surface, there is some homotopically nontrivial
closed curve of minimal length $L$.  Next, recall that
any boost $\gamma \in
\Gamma$ preserves exactly one geodesic on the hyperboloid.  When the
hyperboloid is embedded in the auxiliary Minkowski space, this is
just the geodesic in which the hyperboloid intersects the plane of the
boost $\gamma$.  Let $z$ be any point on this invariant geodesic for
$\gamma$.  The boost $\gamma$ maps $z$ to another point $\gamma z$ whose
distance from $z$ is just the boost parameter $\lambda_\gamma$.  But, when
projected to the Riemann surface, the geodesic segment connecting 
$z$ and $\gamma z$ becomes a homotopically nontrivial closed curve
of length $\lambda_\gamma$.  Thus, $\lambda_\gamma \ge L$. QED

\medskip

{\bf Lemma 2:  
There is a continuous
function $\delta : {\bf R}^+ \rightarrow {\bf R}^+$ such that, 
given any hyperbolic M\"obius transformation $\gamma$ 
on $S^1$
with boost parameter greater than or equal to $L$, 
if the distance between some point $a$ and its image $\gamma a$ is less
than $\epsilon$ then $\gamma$ has a fixed point within a distance
$\delta(\epsilon)$ of $a$.  Furthermore, $\lim_{\epsilon \rightarrow
0}\delta = 0$.}

\medskip

Recall that the set of hyperbolic M\"obius transformations on the circle,
together with the set of parabolic M\"obius transformations on the circle, 
can be identified with the 
closed set $I^0 = \{x:x^2 \ge 0\}$ in 2+1
Minkowski space (i.e., those points spacelike and null separated from the 
origin).
The 
transformations with boost parameter
greater than or equal to $L$ 
correspond to the points 
outside of some timelike hyperboloid.  We may therefore choose
an open set $U$ in the Minkowski space
such that 1) the closure of $U$ is compact, 2) $U$ contains the origin, and
3) any transformation $\gamma$ with $\lambda_\gamma \ge L$ 
is associated with a vector $v_\gamma$
in the Minkowski space that is {\it not} in $U$.

Let $K$ be the intersection of the closure of $U$ with 
$I^0$.  We see that $K$ is compact, and that any $v_\gamma$
is proportional to some element $k_\gamma$ in $K$ (with coefficient greater
than or equal to one).  Note that our lemma will follow
if we can prove the existence
of the required function $\delta$ for all vectors in $K$, since if $\gamma$
moves some point $a$ a distance less than $\epsilon$, so will $k_\gamma$.
Thus, $k_\gamma$ will have a fixed point within $\delta(\epsilon)$ of $a$, 
but any fixed point of $k_\gamma$ is also a fixed point of $\gamma$.

Now, it is clear that, given any transformation $\alpha$ associated
with a vector in $K$, there
is a continuous
function $\delta_\alpha : {\bf R}^+ \rightarrow {\bf R}^+$ with $\lim_{\epsilon
\rightarrow 0}\delta(\epsilon)=0$
such that
if the distance between some point $a$ and its image $\alpha a$ is less
than $\epsilon$, then $a$ is within $\delta_\alpha(\epsilon)$ of a 
fixed point of
$\alpha$.  This is true regardless of whether $\alpha$ is parabolic or 
hyperbolic, and $\delta_\alpha(\epsilon)$ may be taken to depend 
continuously on $(\alpha,\epsilon) \in K \times ({\bf R}^+
\cup \{0\})$.
We may thus use the compactness of $K$ to define $\delta(\epsilon)
= \max_\alpha \delta_\alpha(\epsilon)$, which will be continuous in
$\epsilon$ on ${\bf R}^+ \cup \{0\}$, map ${\bf R^+}$ to ${\bf R^+}$,
and satisfy $\delta(0)=0$. QED

\medskip

These two Lemmas will now allow us to prove the following theorem:

\medskip
{\bf Theorem: Let $\Gamma$ be the group that compactifies the 
hyperboloid to any $g \ge 2$ surface.  Then the fixed points of the action
of $\Gamma$ on $S^1$ are dense.}

\medskip

To complete the proof, consider any geodesic
through the origin of the hyperboloid. This geodesic intersects some
collection of copies $R_\gamma$
of the fundamental region $R_0$, with $R_\gamma$
the image of $R_0$
under the boost $\gamma$ in $\Gamma$.  The action
of this collection 
of boosts on the original geodesic gives a collection
of geodesics ${\cal G}$, the elements of which can be obtained by
moving the segment of the
original geodesic inside $R_\gamma$ back to $R_0$ by means
of the boost $\gamma^{-1}$
and then extending the geodesic to infinity.  The ergodicity results of
\cite{BV} mean 
that the collection ${\cal G}$ is dense
in the set of geodesics on the hyperboloid which pass through $R_0$.
It follows that, given any two geodesics $g_1,g_2$ which pass through 
$R_0$, there are geodesics $g_3$ (close
to $g_1$) and $g_4$ (close to $g_2$) that
can be mapped onto each other
using some boost $\gamma \in \Gamma$.  In particular, this is true of
the endpoints of $g_3,g_4$.

Suppose then that we wish to show the existence of an element $\gamma$
of $\Gamma$ which has a fixed point in
some open interval $(\theta_0,\theta_1)$ along the boundary circle at infinity.
We choose any $\theta \in (\theta_0,\theta_1)$ and connect
the corresponding point at infinity with 
the origin by a geodesic $g_1$ (whose other
endpoint will be at $-\theta$).  Furthermore, 
we can generate a second geodesic $g_2$ which passes through $R_0$
and has endpoints in $(\theta_0, \theta_1)$, 
$(-\theta_1, -\theta_0)$ by acting on this geodesic with a small boost
(which need not lie in $\Gamma$) whose fixed points on the circle
at infinity are, say, $\theta_0$ and $- \theta_1$.  This pair of
geodesics may be drawn as below, and the endpoints are labeled
$a_1,b_1$ and $a_2,b_2$ as shown below.

\centerline{
\epsfbox{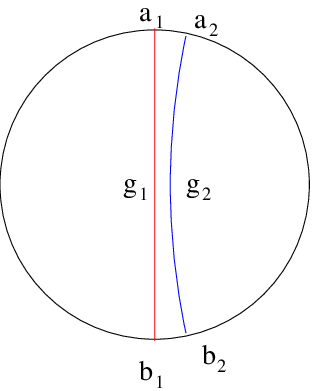}}

\medskip
\centerline{Fig. 4. The geodesics $g_1$ and $g_2$ and their endpoints.}
\medskip

 From our discussion above, we may find two other geodesics $g_3$ and 
$g_4$, arbitrarily close to $g_1$ and $g_2$ respectively, such that
$g_3$ is mapped onto $g_4$
by some boost $B$ in $\Gamma$.  Let us suppose that the endpoints
of $g_3$ and $g_4$ are labeled $a_3,b_3$ and $a_4,b_4$ with
$a_3,a_4 \in (\theta_0,\theta_1)$ and $b_3,b_4 \in (-\theta_1,-\theta_0)$.
Note that we may in fact choose $g_3$, $g_4$, and $B$
so that $B$ maps $a_3$ to $a_4$ and maps $b_3$ to $b_4$.  If our
original choice of $g_3,g_4$ does not allow such a $B$,
then we need only
choose some other geodesic $g_5$ between $g_3$ and $g_4$
and it will be related to either $g_3$ or $g_4$ in the required way.    

Finally, note that there is in fact a choice of $g_3$ such that we may
leave $g_3$ fixed and, by changing our 
choice of $g_4$, make the distance $\epsilon$ between $a_3$ and $a_4$
as small as we like.  Thus, we can arrange $\delta(\epsilon)$ from 
Lemma 2 to be less than the distance between $a_3$ and either of 
$\theta_0$, $\theta_1$.  Lemma 2 then
shows that the associated boost
must have a fixed point in the interval $(\theta_0,\theta_1)$. QED

\end{document}